Title: Reaction Rate Assessment of Multiphonon Relaxation
Authors: Mladen Georgiev (1), Alexander Gochev (2) ((1) Institute of Solid State
Physics, Bulgarian Academy of Sciences, Sofia, Bulgaria, (2) PGA Solutions,
Galloway OH, U.S.A.)
Comments: 30 pages, 9 figures; all pdf format; paper #2 presented at the 2004 Internet
Electronic Conference of Molecular Design

An investigation is made into whether the reaction-rate formalism also applies to describing inelastic-tunneling vibronic processes associated with the absorption and emission of phonons. We propose that Bardeen-Christov's quantum-mechanical approach to the transition (tunneling) probabilities originally intended for elastic tunneling alone can formally be extended to cover inelastic tunneling as well. Namely, we show that inelastic tunneling through the absorption of $p$ phonons can be regarded as an exothermic reaction with zero-point reaction heat $\Theta_p = -ph\nu$. Alternatively, inelastic tunneling through the emission of $p$ phonons will be equivalent to an endothermic reaction with zero-point reaction heat $\Theta_p = +ph\nu$. An analysis made indicates that the transition (tunneling) probabilities expand into Franck-Condon factors revealing the interconnection between reaction-rate and multiphonon relaxation rates. Our proposal confirms to a good approximation the traditional multiphonon fingerprint that the relaxation rate of 1-phonon processes increases linearly with the temperature $T$ at low $T$. The combined reaction rate theory compares favorably with numerical calculations on specific experimental systems.

1. Rationale

One of the essential problems not yet solved satisfactorily by solid state theory is the rate of phonon-coupled processes. There are two large groups: local processes and transport phenomena. The former ones occur in the neighborhood of a site, while the latter transfer charge or matter across the crystal. Examples for local processes are provided by the nonradiative deexcitation of color centers, the local rotation of off-center impurity ions, and, possibly, the generation of lattice defects under ionizing radiation. Examples for transport processes are provided by the diffusion of impurities and the electrical conductivity. There are two main approaches to phonon-coupled processes: the *reaction rate* (RRT) and the *multiphonon* (MPT) theories. Now, while the rate of local processes has been found tractable by the reaction rate theory, the transport rates are more adaptable to multiphonon considerations.

---

The reaction rate theory is based on an occurrence-probability approach in which the transition rate obtains as the sum of elastic-tunneling energy-conserving transitions at the various quantized vibronic energy levels [1-3]. The reaction rate conserves the number of phonons during a horizontal transition, while the multiphonon rate based on the Golden Rule incorporates inelastic-tunneling phonon absorption and emission processes which do not conserve the phonon number [4,5]. The reaction rate accounts for coherent processes, the multiphonon rate is composed of both coherent and incoherent components.

The basic fingerprint of coherency (when the phase of the wave function is retained over distances within the transition range) is the low-temperature reaction rate which is insensitive to the temperature before any essential transitions to the first excited state are initiated by thermal aggitation. In contrast, the benchmark of incoherency (if the phase of the wave function is not retained even at close range) is the low-temperature rate which increases though slowly with the temperature within the same range where the reaction rate remains constant.

It is not immediately clear just why the translational motion does require incoherency, while the local rotation which imitates migration does not. Whatever the reason, we want to cast a bridge between the two approaches and compare them to specific experimental situations. Emphasis will be laid on transport situations since they require both coherent and incoherent rate components. The basic advantage of using the reaction rate is that the method allows for describing the whole temperature dependence as a *continuous* rate, incorporating the low-temperature tunneling range, followed by the transition range, or activated-tunneling range, where excited vibronic states start filling up, and finally by the thermally-activated classical range.

## 2. Basic equations

### 2.1. Nuclear Tunneling

Following Christov, the nuclear- (configurational-) tunneling probability $W_{conf}(E_n,E_n)$ for a horizontal isoenergetic transition conserving the phonon number can be calculated using the extension of a formula due to Bardeen [2,3]

$$W_{conf}(E_n) = 4\pi^2 |U_{if}(E_n)|^2 \sigma_i(E_n)\sigma_f(E_n) \tag{1}$$

where

$$U_{if}(E_n) = -(\hbar^2/2m)\{\chi_i(E_n)[d\chi_f^*(E_n)/dq] - \chi_f(E_n)[d\chi_i^*(E_n)/dq]\}\big|_{q=q_C} \tag{2}$$

is the potential induced by the transition current. Here $\sigma_i(E_n)$ and $\sigma_f(E_n)$ are the DOS, $\chi_i(E_n)$, $\chi_f(E_n)$ are the nuclear-oscillator wave functions in the initial and final electronic states, respectively, $E_n$ is the energy of the vibronic transition. Using harmonic-oscillator wave functions normalized in Q-space:

$$\chi_n(q)_\pm = [\sqrt{(\alpha/\pi)}/2^n n!]^{1/2} H_n(q)\exp(-(q\pm q_0)^2/2) \tag{3}$$

where $\alpha = M\omega^2/\hbar\omega$, $q = \sqrt{\alpha} Q$ is the scaled and Q the actual configurational coordinate, $q_0$ is the absolute position along q of the well bottom, $q_C$ is the crossover coordinate. $H_n(q)$ are Hermite polynomials, $E_n = (n+½)\hbar\omega$, $\sigma_i(E_n) = \sigma_f(E_n) = 1/\hbar\omega$.

For an inelastic-tunneling transition to a higher-lying state at $E_{n+p}$, we will have instead (cf. Appendix 1)

$$W_{conf}(E_n,E_{n+p}) = 4\pi^2 |U_{if}(E_n,E_{n+p})|^2 \sigma_i(E_n)\sigma_f(E_{n+p}) \tag{4}$$

$$U_{if}(E_n, E_{n+p}) = -(\hbar^2/2M)\{ u_i(E_n) [du_f^*(E_{n+p})/dq] -$$

$$u_f(E_{n+p}) [du_i^*(E_n)/dq]\}|_{q=qC} \quad (5)$$

It is easy to see that $U_{if}(E_n, E_{n+p})$ is formally identical to the current-induced potential of an exothermic process. Indeed, shifting the final-state adiabatic parabola along the vertical energy axis until the vibronic energy levels $E_n$ and $E_{n+p}$ meet each other, $U_{if}(E_n, E_{n+p})$ is seen characteristic of an exothermic process whose zero-point reaction heat is $\Theta_p = -p\hbar\omega$. What we do is a vibronic potential energy transform of the form $V(q)' = V(q) + \Theta_p$. This is illustrated in Figure 1 depicting adiabatic potential energies and the arrows of inelastic tunneling transitions on the left and their elastic-tunneling equivalents on the right. But, it should be stressed that as the vertical shifts of a parabola preserve the positions of the well bottoms, the suggested multiphonon transitions should occur at constant $q_0$, independent of $p$. No similar conclusion can be drawn as regards the crossover coordinate $q_C=0$. Alternatively, one may keep $q_C$ independent of and move $q_0$ in concert with $p$. Other choises of independent quantities are conceivable too.

Consequently, an elastic-tunneling exothermic process with $\Theta_p = -p\hbar\omega$ is equivalent to the absorption of $p$ phonons by the vibronic system which first gets excited and then transfers them to the thermal bath. An endothermic process with $\Theta_p = +p\hbar\omega$ is equivalent to the emission of $p$ phonons which first deexcite the vibronic system and are then given back by the bath.

The elastic nuclear-tunneling probability at any finite $\Theta_p$, nonpositive or positive, reads:

$$W_{conf}(E_n, E_m) = \pi\{[F_{nm}(\xi_0, \xi_C)]^2/2^{n+m} n!m!\} \exp(-\varepsilon_R/\hbar\omega) \exp(-\Theta_p^2/\hbar\omega\varepsilon_R),$$

where the vibronic level number in final electronic state is $m = n + p$ (at $\Theta_p < 0$) and $m = n - p$ (at $\Theta_p > 0$), or equivalently,

$$W_{conf}(E_n, \Theta_p) = \pi\{[F_{n,n\pm p}(\xi_0, \xi_C)]^2/2^{2n\pm p} n!(n\pm p)!\} \exp(-\varepsilon_R/\hbar\omega) \exp(-\Theta_p^2/\hbar\omega\varepsilon_R)$$

$$(6)$$

Using harmonic-oscillator wave functions

$$F_{nm}(\xi_0, \xi_C) = \xi_0 H_n(\xi_C) H_m(\xi_C - \xi_0) - 2n H_{n-1}(\xi_C) H_m(\xi_C - \xi_0) +$$

$$2m H_n(\xi_C) H_{m-1}(\xi_C - \xi_0), \quad (7)$$

or equivalently,

$$F_{n,n\pm p}(\xi_0, \xi_C) = \xi_0 H_n(\xi_C) H_{n\pm p}(\xi_C - \xi_0) - 2n H_{n-1}(\xi_C) H_{n\pm p}(\xi_C - \xi_0) +$$

$$2(n\pm p) H_n(\xi_C) H_{n\pm p-1}(\xi_C - \xi_0) \quad (7')$$

Here

$$\varepsilon_C \equiv \tfrac{1}{2} K Q_C^2 = \tfrac{1}{2} \hbar\omega\, q_C^2 = (\varepsilon_R + \Theta_p)^2 / 4\varepsilon_R \qquad (8)$$

is the crossover energy,

$$\varepsilon_R \equiv 2 \times \tfrac{1}{2} KQ_0^2 = KQ_0^2 = \hbar\omega\, q_0^2 \qquad (9)$$

is the lattice reorganization energy, K is the stiffness. In so far as $q_0$ is p-independent, so is $\varepsilon_R$. $V_{12} = \tfrac{1}{2}\varepsilon_{gap} = 2\eta\varepsilon_{JT}$ is the crossover resonance half- splitting energy.

In order to avoid complications arising from the electron-transfer terms to be discussed shortly, we consider at this point the rate of an *adiabatic* process in which the electronic state changes with certainty at all transfer energies. With the above considerations in mind, we rewrite the reaction rate incorporating both the isophonon elastic-tunneling rate and the multiphonon inelastic-tunneling corrections in the form:

$$\mathfrak{R}_R(T) = 2\nu\sinh(\hbar\omega/2k_BT)\{\sum_{n=0}^{\infty} W_{conf}(E_n,E_n) \exp(-E_n/k_BT)\big|_{\Theta_p=0} +$$

$$\sum_{p=1}^{\infty} W_{conf}(E_0,E_p) \exp(-E_0/k_BT)\big|_{\Theta_p<0} +$$

$$\sum_{p=1}^{\infty} \sum_{n=1}^{\infty} W_{conf}(E_n,E_{n+p}) \exp(-E_n/k_BT)\big|_{\Theta_p<0} +$$

$$\sum_{p=1}^{\infty} \sum_{n=1}^{\infty} W_{conf}(E_n,E_{n-p}) \exp(-E_n/k_BT)\big|_{\Theta_p>0} \} \qquad (10)$$

in which the underlying probability terms should be understood as ones controlling the corresponding exo- or endo-thermic processes at increasing reaction heats $|\Theta_p|$ and at $\Theta_p$-independent well bottom coordinate $q_0$. At $p=0$ the n-sum on the first line of eq. (10) corresponds to the reaction rate of an elastic isothermic process. The *p*-sum at n=0 on the second line is the rate of multiphonon absorption starting from the ground-state vibronic energy level (n=0). At $p\geq 1$, the n-sum on the third line gives the contribution of exothermic (*phonon absorption*) processes ($\Theta_p<0$) starting from an excited state vibronic energy level (n≥1). Finally, the sum on the fourth line gives the contribution of endothermic (*phonon emission*) processes ($\Theta_p>0$). In so far as there could be no phonon emission from the ground-state vibronic energy level (n=0), all emission transitions should start from n≥1. Both the individual exo- or endo- thermic processes are equivalent to the inelastic tunneling steps. It is clear from eq. (6) that it is the probability terms at small *p* which should contribute most to the multiphonon corrections.

Considering that $E_n = (n + \tfrac{1}{2})\hbar\omega$ for $n \geq 0$, we can take $\exp(-E_0/k_BT) = \exp(-\hbar\omega/2k_BT)$ out of the sum in (10) to obtain an equivalent form of $\mathfrak{R}_R(T)$:

$$\mathfrak{R}_R(T) = \nu\,[1- \exp(-\hbar\omega/k_BT)]\{ \sum_{n=0}^{\infty} W_{conf}(E_n,E_n) \exp(-n\hbar\omega/k_BT)\big|_{\Theta_p=0} +$$

$$\sum_{p=1}^{\infty} W_{conf}(E_0,E_p) \big|_{\Theta_p<0} +$$

$$\sum_{p=1}^{\infty} \sum_{n=1}^{\infty} W_{conf}(E_n,E_{n+p}) \exp(-n\hbar\omega/k_BT) \big|_{\Theta_p<0} +$$

$$\sum_{p=1}^{\infty} \sum_{n=1}^{\infty} W_{conf}(E_n,E_{n-p}) \exp(-n\hbar\omega/k_BT) \big|_{\Theta_p>0} \} \tag{11}$$

Here the nuclear tunneling probabilities, e.g. $W_{conf}(E_0,E_p)$, depend on $p$ through the $p$-dependencies of the Hermite polynomials in equation (7) via eq. (8), while eq. (9) is $p$-independent.

From equations (6) and (7) we get for the second-line term:

$$\sum_{p=1}^{\infty} W_{conf}(E_0,E_p) = \pi\exp(-\varepsilon_R/\hbar\omega) \sum_{p=1}^{\infty} \{[F_{0p}(\xi_0,\xi_C)]^2/2^p p!\}\exp(-p^2\hbar\omega/\varepsilon_R)$$

where

$$F_{0p}(\xi_0,\xi_C) = \xi_0 H_p(\xi_C-\xi_0) + 2p H_{p-1}(\xi_C-\xi_0)$$

The main contribution to the rate is from one-phonon absorption processes. We get for $p=1$:

$$W_{conf}(E_0,E_1) = \pi\exp(-\varepsilon_R/\hbar\omega) \{[F_{01}(\xi_0,\xi_C)]^2/2\} \exp(-\hbar\omega/\varepsilon_R)$$

with

$$F_{01}(\xi_0,\xi_C) = 2[\xi_0(\xi_C-\xi_0) + 1].$$

### 2.2. Electron-transfer probability

In an elastic-tunneling transition at any $p$, given the nuclear-tunneling probability $W_{conf}(E_n,E_{n\pm p}) \equiv W_{conf}(E_n,\Theta_p)$, the overall transfer probability at $E_n$ obtains as a product of $W_{conf}(E_n,\Theta_p)$ times the probability for a change of the electronic state $W_{el}(E_n,\Theta_p)$, viz.

$$W(E_n,\Theta_p) = W_{conf}(E_n,\Theta_p) W_{el}(E_n,\Theta_p). \tag{12}$$

$W_{el}(E_n,\Theta_p)$, regarded as Landau-Zener's electron-transfer probability, is [3,6]

$$W_{el\ underbarrier}(E_n,\Theta_p) = 2\pi\gamma^{2\gamma-1} \exp(-2\gamma) / [\Gamma(\gamma)]^2$$

$$W_{el\ overbarrier}(E_n,\Theta_p) = 2\{[1 - \exp(-2\pi\gamma)] / [2 - \exp(-2\pi\gamma)]\} \tag{13}$$

where $\gamma(E_n,\Theta_p)$ is Landau-Zener's parameter

$$\gamma(E_n,\Theta_p) = (V_{12}^2/2\hbar\omega)[\varepsilon_R|E_n-\varepsilon_{Cp}|]^{-1/2} \tag{14}$$

A transfer process is termed *adiabatic* if all $W_{el}(E_n, E_{n\pm p}) = 1$. Otherwise it is termed *not-adiabatic*, with the suffix changing to *non-* to mark the extreme case of $W_{el}(E_n, E_{n\pm p}) \ll 1$.

In view of the observed equivalence between the inelastic-tunneling transitions in an isothermic situation and the elastic-tunneling transitions in an exo- or endo- thermic situation, we assume that the electron-transfer term $W_{el}(E_n, E_{n\pm p})$ applies to a horizontal transition at initial-state energy $E_n$ which corresponds to the energy $E_{n\pm p} = E_n \pm p\hbar\omega$ in final state. In other words, the electronic resonance transitions should occur under the conditions covered by Landau-Zener's theory [6].

### 3. Comparison between multiphonon rates and reaction rates

Based on GR, we define the fundamental static-basis multi-phonon rate as [7]:

$$\Re_{i \to f} = (2\pi/\hbar)\, 2\sinh(\hbar\omega/2k_BT) \sum_{n'=0}^{\infty} \sum_{n=0}^{\infty} \exp(-\hbar\omega(n+\tfrac{1}{2})/k_BT) \times$$

$$\left| F_{12} \int dQ\, \chi_{n'}(Q+Q_0)\, Q\, \chi_n(Q-Q_0) \right|^2 \delta(J_f + E_{n'} - J_i - E_n) \qquad (15)$$

where $F_{12}$ is the linear electron-mode coupling coefficient, $J_i$, and $J_f$ are the electron binding energies, $E_n = (n+\tfrac{1}{2})\hbar\omega$, is the vibrational mode energy to harmonic approximation, $\pm q_0$ are the positions of the lateral-well minima along the mode coordinate q. Due to the finite overlap of harmonic oscillator eigenfunctions of displaced oscillators, the integral in (15) is significant in the vicinity of the crossover point $q_C$. For this reason we write to within Condon's approximation

$$\Re_{i \to f} = (2\pi/\hbar)\, 2\sinh(\hbar\omega/2k_BT)\, |F_{12}Q_C|^2 \sum_{n'=0}^{\infty} \sum_{n=0}^{\infty} \exp(-\hbar\omega(n+\tfrac{1}{2})/k_BT) \times$$

$$\left| \langle n'; i(-q_0) | n; f(+q_0) \rangle \right|^2 \delta(J_f + E_{n'} - J_i - E_n) \qquad (16)$$

At this point we also remind that

$$F_{12}\, Q_C = \langle \psi_f(x) | H_{int} | \psi_i(x) \rangle \qquad (17)$$

For symmetric wells we assume $J_i = J_f$. We shall exemplify both similarities and differences between reaction rates and 1-phonon rates which are the main multiphonon components; now, $n' = n+1$. For this case

$$\Re_{i \to f} = \nu\, [1 - \exp(-\hbar\omega/k_BT)] \sum_{n=0}^{\infty} \exp(-n\hbar\omega/k_BT) \times$$

$$\left| \langle n+1; i(-q_0) | n; f(+q_0) \rangle \right|^2 |F_{12}\, Q_C|^2\, (4\pi^2/\hbar)\, \delta(J_f + E_{n+1} - J_i - E_n) \qquad (18)$$

The 1-phonon rate is to be compared with the reaction rate of a non-adiabatic process:

$$\Re_R = \nu\, [1 - \exp(-\hbar\omega/k_BT)] \left\{ \sum_{n=0}^{\infty} W_{conf}(E_n, E_n) \exp(-n\hbar\omega/k_BT) \Big|_{\Theta p=0} + \right.$$

$$W_{conf}(E_0,E_1) \mid_{\Theta 1<0} + \sum_{n=1}^{\infty} W_{conf}(E_n,E_{n+1}) \exp(-n\hbar\omega/k_BT) \mid_{\Theta p<0} +$$

$$\sum_{n=1}^{\infty} W_{conf}(E_n,E_{n-1}) \exp(-n\hbar\omega/k_BT) \mid_{\Theta p>0} \} (4\pi V_{12}^2/\hbar\omega E_R) \qquad (19)$$

where according to eq's (4), (5), and (14):

$$W_{conf}(E_n,E_{n+1}) = (\pi\hbar^2/M)^2 \mid \chi_{i\ n}(q+q_0)d\chi_{f\ n+1}^*(q-q_0)/dq -$$

$$\chi_{f\ n+1}(q-q_0)d\chi_{i\ n}^*(q+q_0)/dq \mid^2_{q=qC}$$

$$W_{conf}(E_n,E_{n-1}) = (\pi\hbar^2/M)^2 \mid \chi_{i\ n}(q+q_0)d\chi_{f\ n-1}^*(q-q_0)/dq -$$

$$\chi_{f\ n-1}(q-q_0)d\chi_{i\ n}^*(q+q_0)/dq \mid^2_{q=qC}$$

$$W_e \sim 4\pi\gamma_n = 4\pi(V_{12}^2/2\hbar\omega)/\sqrt{(E_R \mid E_C-E_n \mid)} \sim 4\pi V_{12}^2/\hbar\omega E_R \text{ for n small}$$

Comparing (18) with (19) we see that $F_{12}Q_C = V_{12}$ and remind that $q = \sqrt{\alpha}Q$ with $\alpha = M\omega/\hbar$. The oscillator wave functions read as in eq. (3):

$$\chi_{i/f\ n}(q \pm q_0) = N_n H_n(q \pm q_0)\exp(-(q \pm q_0)^2/2), \quad N_n = [\sqrt{(\alpha/\pi)}/2^n n!]^{1/2}$$

We further make use of the recursion relations

$$H_{n+1}(q) - 2qH_n(q) + 2nH_{n-1}(q) = 0$$

$$H_n'(q) \equiv dH_n(q)/dq = 2nH_{n-1}(q)$$

and insert the latter into the wavefunction derivatives to get

$$d\chi_{i/f\ n}(q \pm q_0)/dq = \chi_{i/f\ n-1}(q \pm q_0) - (q \pm q_0)\chi_{i/f\ n}(q \pm q_0)$$

Now, putting the result into the nuclear terms we obtain

$$W_{conf}(E_n,E_{n+1}) = (\pi\hbar^2/M)^2 \mid \chi_{i\ n}(q+q_0) [\chi_{f\ n}(q-q_0) - (q-q_0)\chi_{f\ n+1}(q-q_0)]^* -$$

$$\chi_{f\ n+1}(q-q_0) [\chi_{i\ n-1}(q+q_0) - (q+q_0)\chi_{i\ n}(q+q_0)]^* \mid^2_{q=qC}$$

$$W_{conf}(E_n,E_{n-1}) = (\pi\hbar^2/M)^2 \mid \chi_{i\ n}(q+q_0) [\chi_{f\ n-2}(q-q_0) - (q-q_0)\chi_{f\ n-1}(q-q_0)]^* -$$

$$\chi_{f\ n-1}(q-q_0) [\chi_{i\ n-1}(q+q_0) - (q+q_0)\chi_{i\ n}(q+q_0)]^* \mid^2_{q=qC}$$

$$W_{conf}(E_n,E_n) = (\pi\hbar^2/M)^2 \mid \chi_{i\ n}(q+q_0) [\chi_{f\ n-1}(q-q_0) - (q-q_0)\chi_{f\ n}(q-q_0)]^* -$$

$$\chi_{f\ n}(q-q_0) [\chi_{i\ n-1}(q+q_0) - (q+q_0)\chi_{i\ n}(q+q_0)]^* \mid^2_{q=qC}$$

or, in equivalent notations,

$$W_{conf}(E_n, E_{n+1}) = (\pi\hbar^2/M)^2 \left| \{ [ <n,f| - (q+q_0) <n+1,f| ] | n,i> - \right.$$

$$\left. [ <n-1,i| - (q+q_0) <n,i| ] | n+1,f> \} \delta(q-q_C) \right|^2$$

$$W_{conf}(E_n, E_{n-1}) = (\pi\hbar^2/M)^2 \left| \{ [ <n-2,f| - (q-q_0) <n-1,f| ] | n,i> - \right.$$

$$\left. [ <n-1,i| - (q+q_0) <n,i| ] | n-1,f> \} \delta(q-q_C) \right|^2$$

$$W_{conf}(E_n, E_n) = (\pi\hbar^2/M)^2 \left| \{ <n-1,f|n,i> - <n-1,i|n,f> + \right.$$

$$\left. (q+q_0) <n,i|n,f> - (q-q_0) <n,f|n,i> \} \delta(q-q_C) \right|^2 \tag{20}$$

Eq's (20) show how the reaction-rate probabilities for inelastic nuclear tunneling expand in terms of multiphonon Franck-Condon factors. The latter comprise 0-phonon and 1-phonon terms mainly, though there appears a 2-phonon ingredient as well, both in absorption and in emission. Even the horizontal transition probabilities are seen to incorporate a small 1-phonon mixture along with the main 0-phonon terms. We realize that only transitions starting from the vibronic ground state at n=0 are absolutely clean of any 1-phonon mixture.

It would be interesting to compare between the zero-point nonadiabatic isothermic rates by the two alternative approaches. The 0-phonon rate at zero point by the multiphonon theory involves transitions between the two ground state vibronic levels at $n = 0$ of the two-site problem and from eq.(16) it is

$$\Re_{i \to f}^0 = (2\pi/\hbar) \left| F_{12} Q_C \right|^2 \left| <0; i(-q_0) | 0; f(+q_0)> \right|^2 \delta(J_f + E_{n'} - J_i - E_n)$$

$$= (2\pi/\hbar) \left| F_{12} Q_C \right|^2 \exp(-\varepsilon_R/\hbar\omega) \, \delta(J_f + E_{n'} - J_i - E_n)$$

Likewise, the zero-point reaction rate from eq.(6) also comprises elastic tunneling transitions between ground state vibronic levels:

$$\Re_R^0 = \nu \; W_{conf}(E_0, E_0) \, W_{el}(E_0, E_0) = (1/2h) \, V_{12}^2 \, \sqrt{(\varepsilon_R/|\varepsilon_C - \tfrac{1}{2}\hbar\omega|)} \, \exp(-\varepsilon_R/\hbar\omega) \, (1/\hbar\omega)$$

$$\sim (2\pi/\hbar) \, V_{12}^2 \, \exp(-\varepsilon_R/\hbar\omega) \, (1/\hbar\omega),$$

the last line holding good at low frequency such that $\varepsilon_C \gg \tfrac{1}{2}\hbar\omega$. We see that $\Re_{i \to f}^0$ and $\Re_R^0$ agree with each other if the delta-function in the former and $1/\hbar\omega$ in the latter represent similar DOS (density of states).

Finally, it may be instructive to evaluate the Franck-Condon integrals appearing in (15) and subsequently. For an illustration we set $n = 0$ in 1-phonon absorption to obtain

directly

$$I_{FC} = \int_{-\infty}^{+\infty} dq\ \chi^*_{f\,n+1}(q-q_0)\ q\ \chi_{i\,n}(q+q_0) = (1/2\sqrt{\pi}) \exp(-q_0^2)$$

The same integral can also be evaluated using Condon's approximation in order to check its viability:

$$I_{FCC} = q_C \int_{-\infty}^{+\infty} dq\ \chi^*_{f\,n+1}(q-q_0)\chi_{i\,n}(q+q_0) = q_C q_0 (1/\sqrt{\pi}) \exp(-q_0^2)$$

The ratio of the two is $I_{FCC} / I_{FC} = 2q_0 q_C$. Condon's approximation is sometimes perhaps too crude.

## 4. Examples for local and transport relaxation data

Among the typical experimental data that lend support to our proposition are the temperature dependences of the local relaxation of excited F centers in KCl and the local rotation of off-center silver ion impurities in RbCl and RbBr, as shown in Fig.2 (a) through (c) [8,9]. Other examples are provided by the local rotation of the off-center $F^-$ impurity in NaBr, KI, and RbI [9].

From the transport data we selected the temperature dependencies of carbon diffusion in $\alpha$-iron and of the axial and in-plane currents in $La_{2-x}Sr_x CuO_4$, as shown in Fig.3 (a) through (c) [10,11].

We see that the main difference between local and transport data is in the character of the low-temperature tunneling branch: largely flat for the former and slowly increasing with T for the latter. This prompted the use of the pure reaction rate to fit the local data, while the transport data have been covered by a mixture of reaction rate corrected for one-phonon absorption below 100 K. The relative weight of inelastic transitions versus thermally-activated horizontal transitions near the bending point $T_t$ has shown that the former no longer make any significant contribution to the basic elastic process.

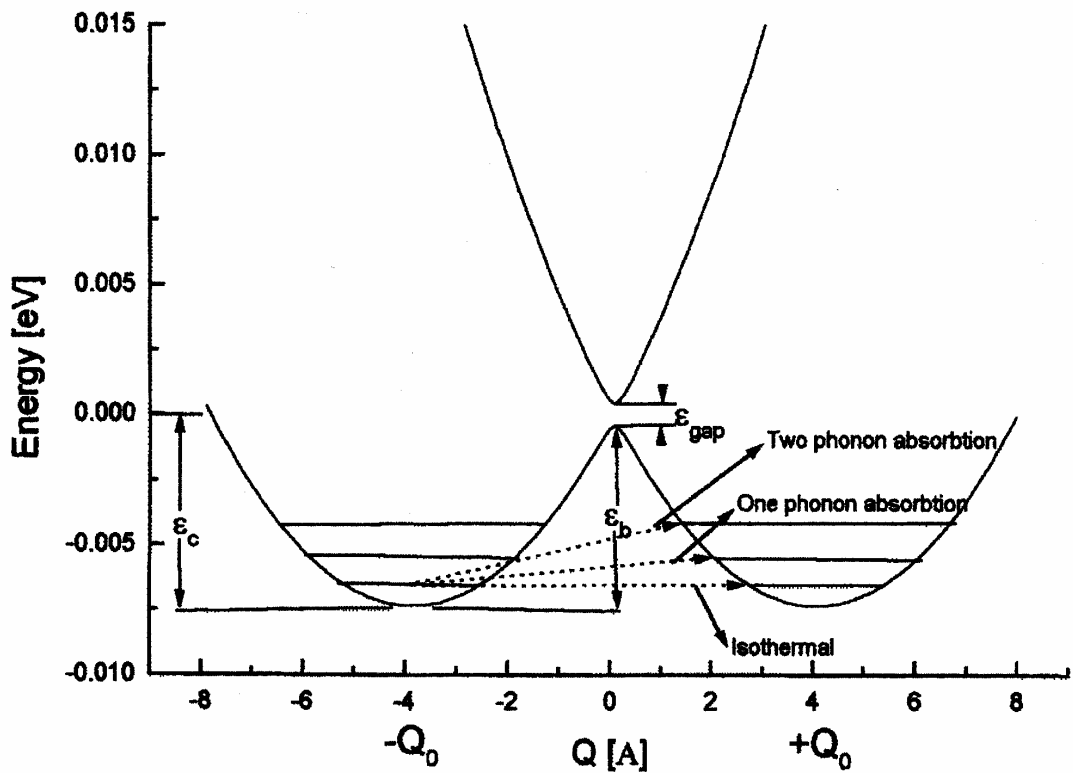

Figure 1 (a)

Figure 1 (a). Illustrating isothermic adiabatic potential energy profiles for tunneling transitions in local and transport situations: horizontal (elastic) tunneling transitions (horizontal arrows), nonhorizontal (inelastic) tunneling transitions (inclined arrows). Q is the coupled mode coordinate.

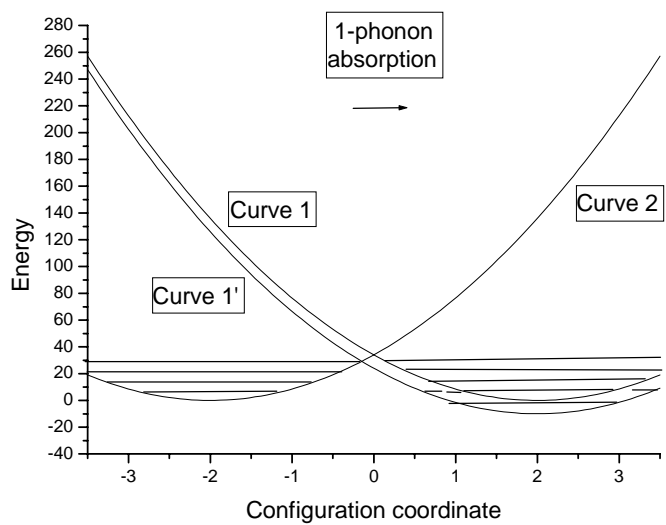

Figure 1 (b)

Figure 1 (b). Illustrating the reaction rate approach to 1-phonon absorption: The Curve 2–Curve 1 pair depicts an isothermic situation by means of two dispaced diabatic potentials. The 1–phonon absorption lifts the system from the lowest level on Curve 2 to the next higher level on Curve 1. The nuclear tunneling probability for the nonhorizontal step is identical to the one for a horizontal step within a Curve 2 – Curve 1' pair in which Curve 1' is displaced downwards along the energy axis by a phonon quantum $-\hbar\omega$. The Curve 2 – Curve 1' pair forms an exothermic doublet. The arrow shows the direction of the process from left to right.

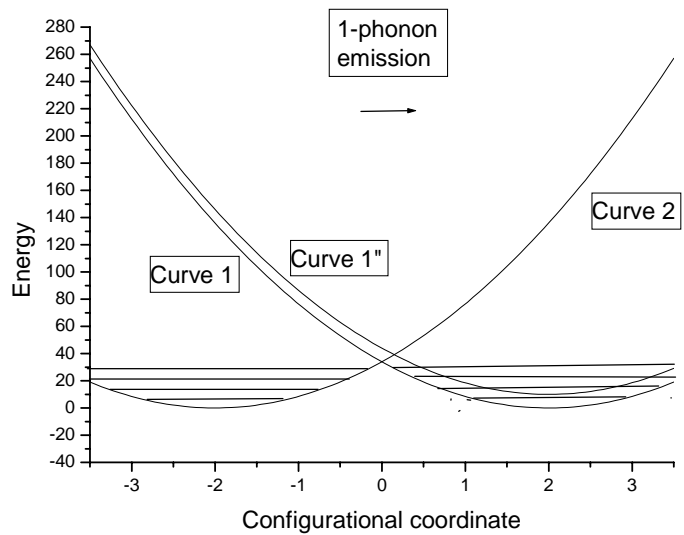

Figure 1 (c)

Figure1 (c). Illustrating the reaction rate approach to 1-phonon emission: The Curve 2–Curve 1 pair depicts an isothermic situation by means of two dispaced diabatic potentials. The 1–phonon emission makes the system dip from the lowest excited level on Curve 2 to the ground state level on Curve 1. The nuclear tunneling probability for the nonhorizontal step is identical to the one for a horizontal step within a Curve 2–Curve 1" pair in which Curve 1" is displaced upwards along the energy axis by one phonon quantum $+\hbar\omega$. The Curve 2 –Curve 1" pair forms an endothermic doublet. The arrow shows the direction of the process from left to right

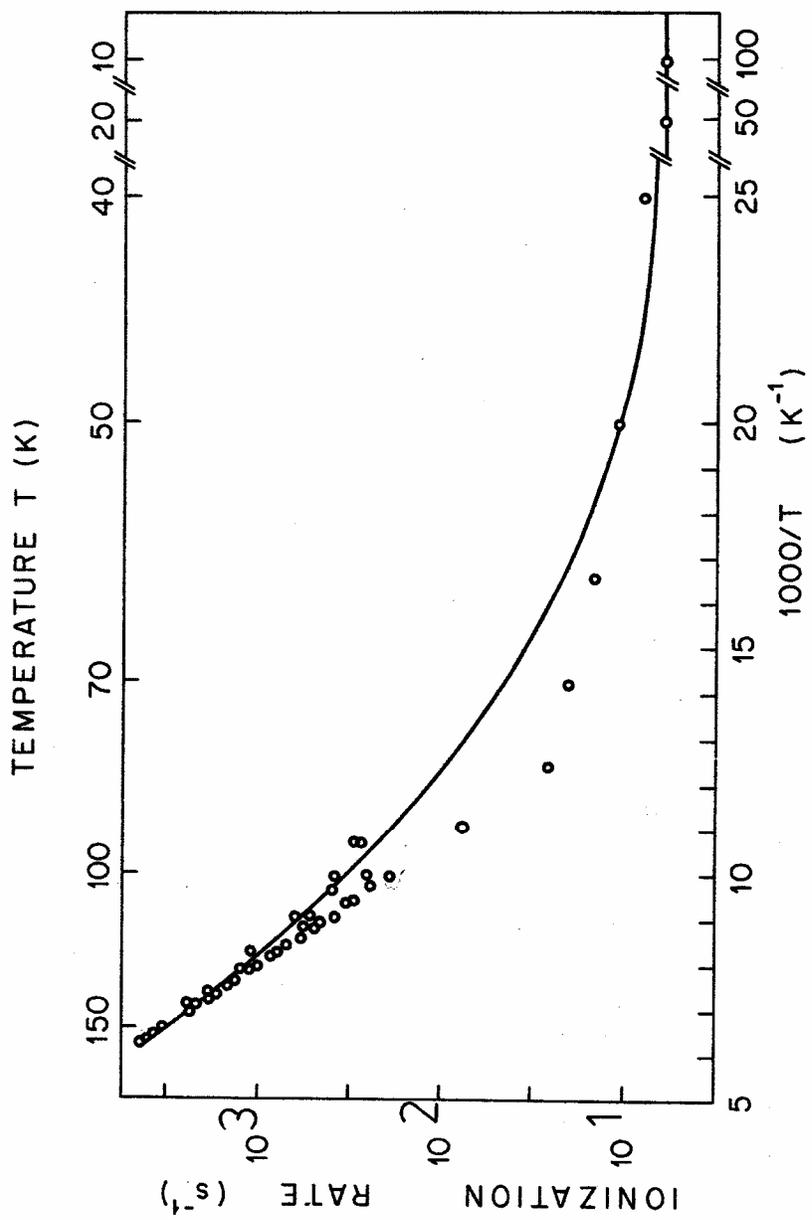

Figure 2 (a)

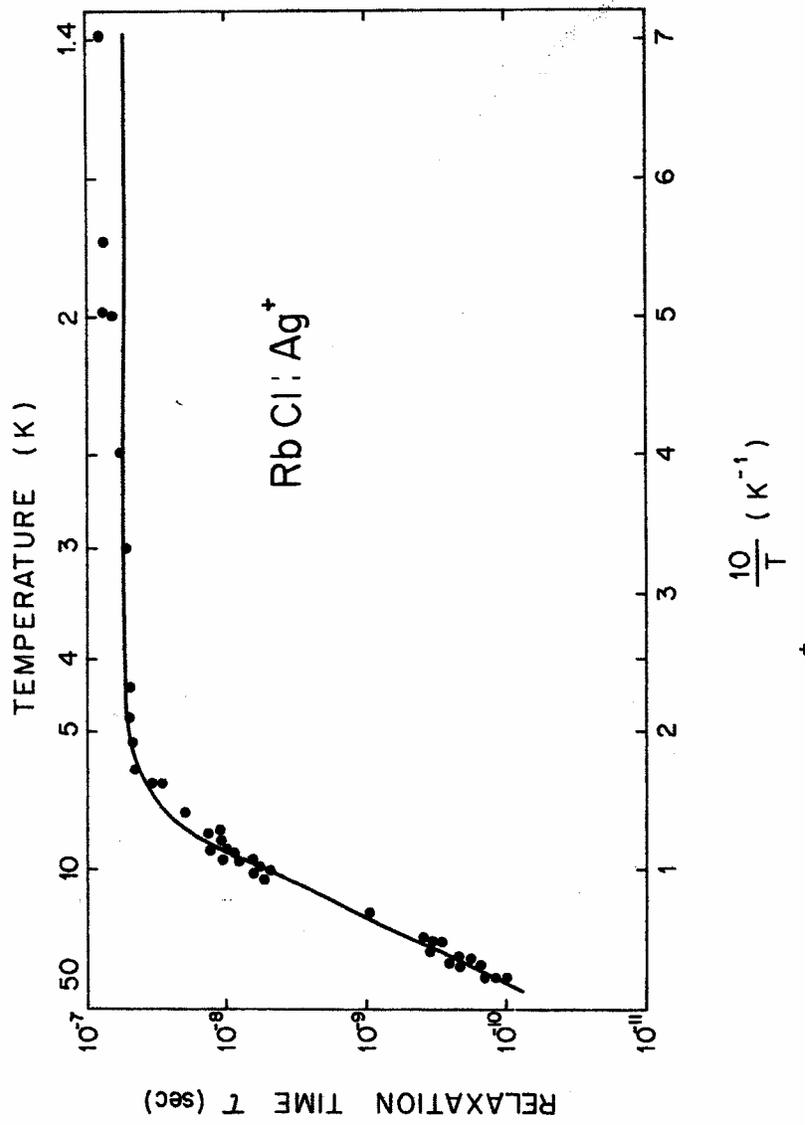

Figure 2 (b)

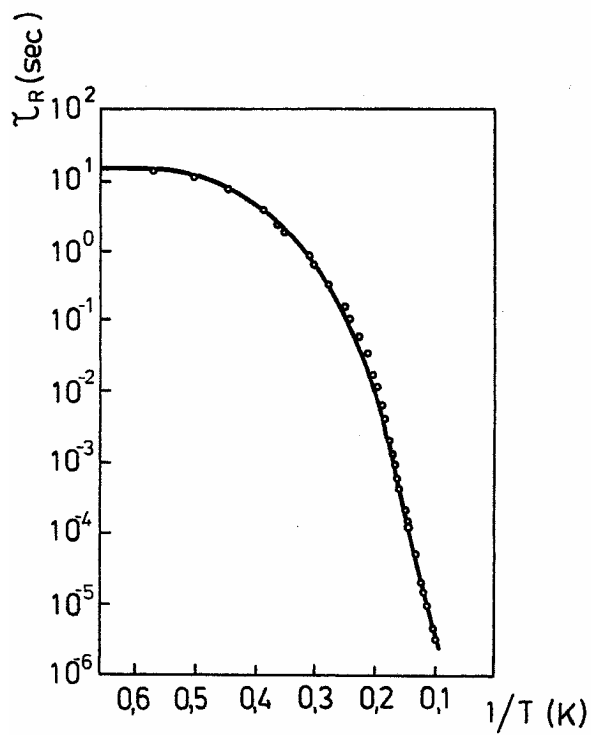

Figure 2 (c)

Figure 2. Examples for the local relaxation of the excited F center in KCl (a) and the local rotation of the $Ag^+$ impurity in RbCl (b) and RbBr (c) tractable by the horizontal (elastic) tunneling reaction rate theory.

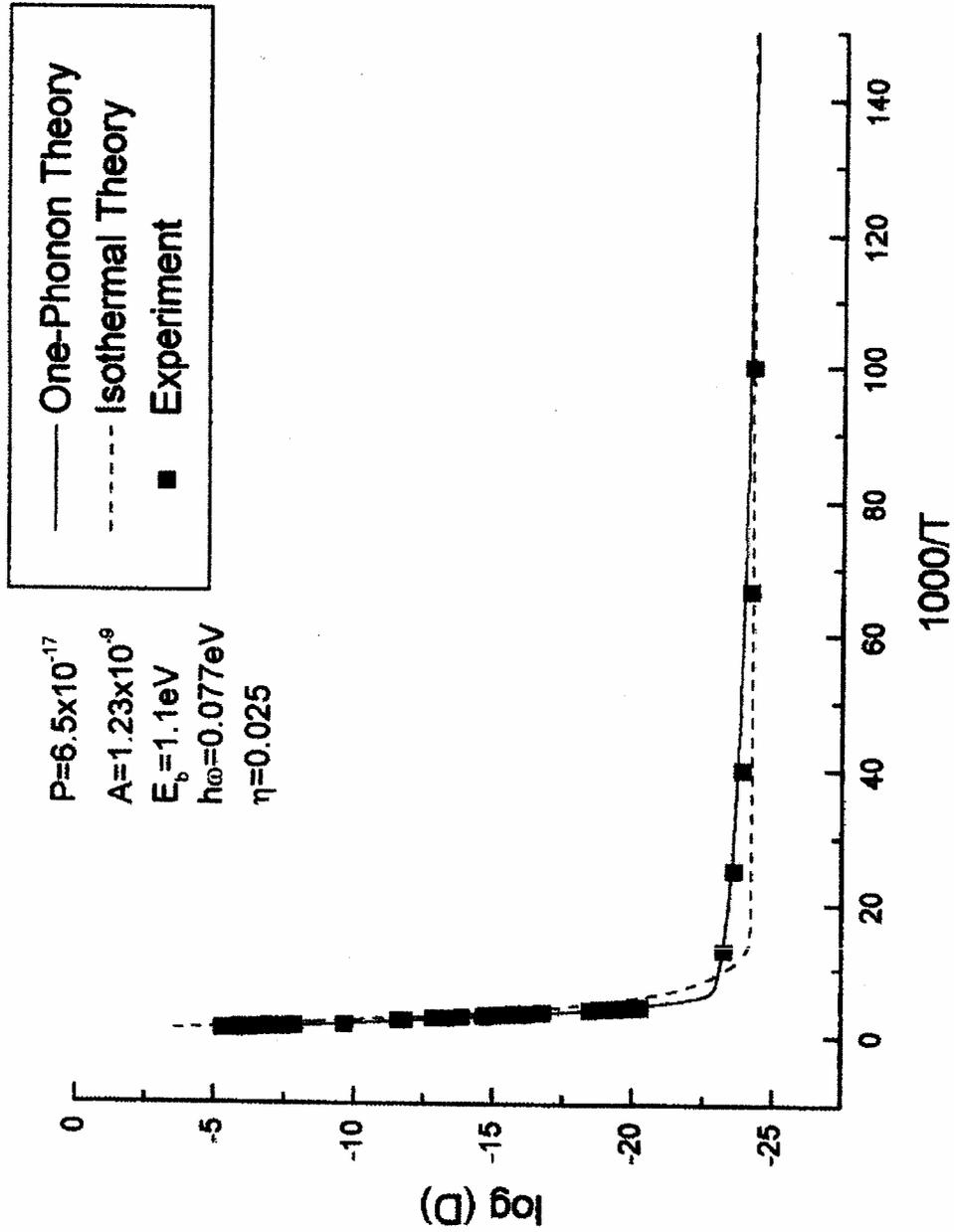

Figure 3 (a)

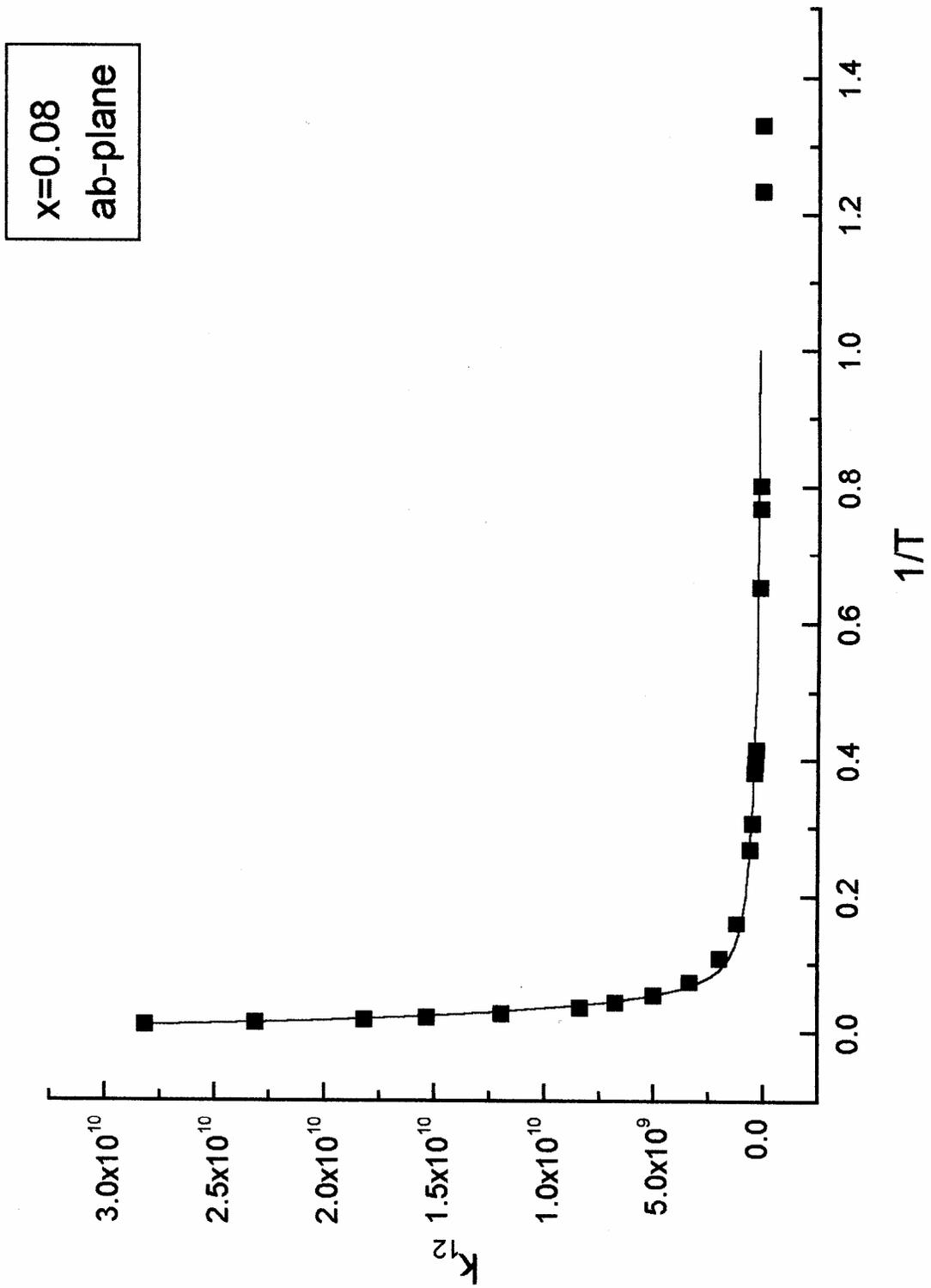

Figure 3 (b)

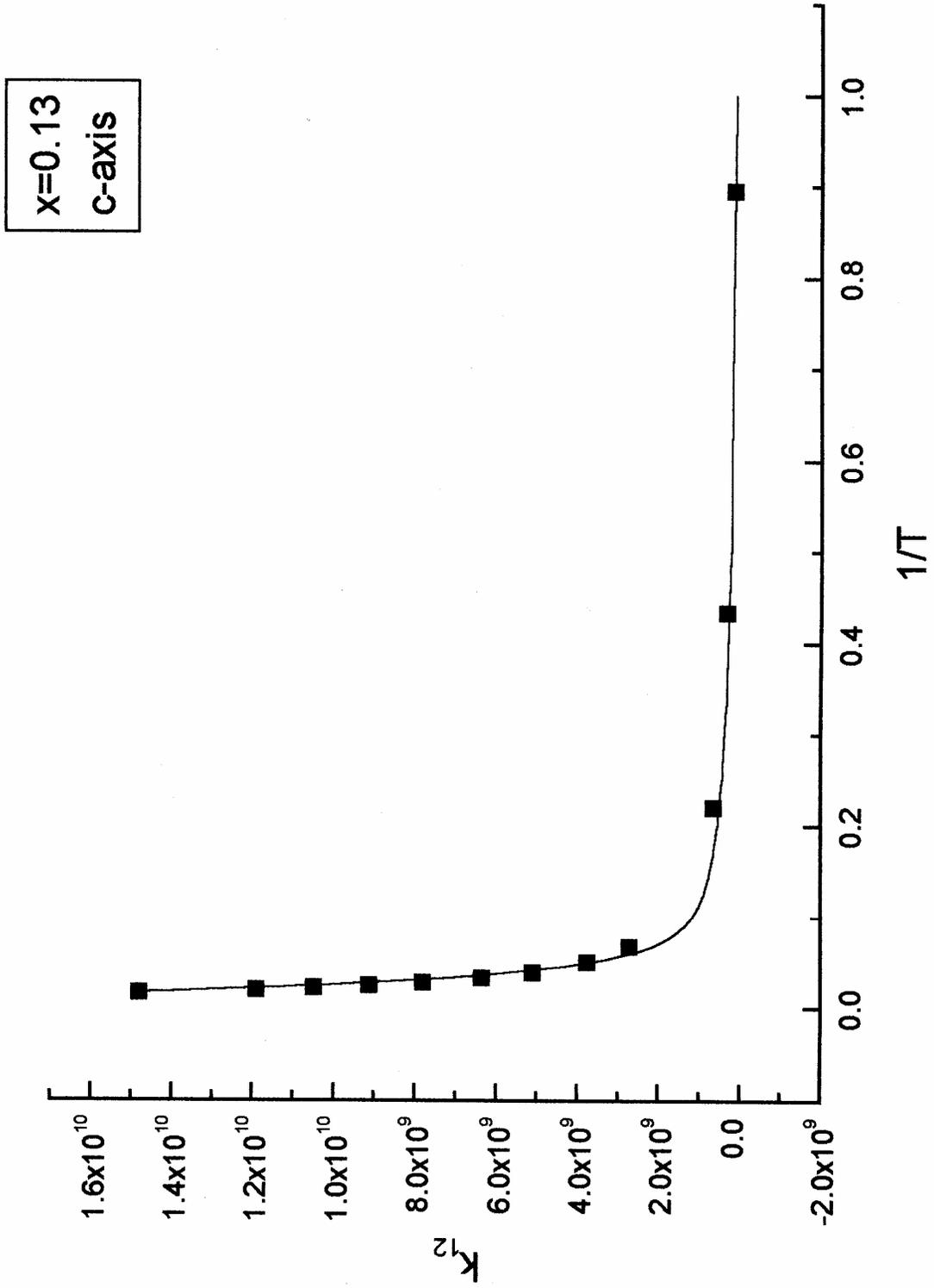

Figure 3 (c)

Figure 3. Examples for fits to experimental transport data (symbols) by the combined reaction rate theory. The data are provided by carbon diffusion in $\alpha$–iron (a) and the axial (b) and in-plane (c) electrical conductivities in $La_{2-x}Sr_xCuO_4$. The solid lines in (a) through (c) are fits by the combined reaction rate incorporating bot the elastic isothermic terms and the $\propto T$ 1-phonon inelastic correction. The dashed line in (a) is a fit obtained by means of the isothermic rate alone.

## 5. Combined low-temperature reaction rate

From eq.(11) we get the combined *lowest-temperatures rate* composed of isothermic and exothermic components, respectively:

$$\Re_R(T)_{lowest\ T} = \nu\ [1-\exp(-\hbar\omega/k_BT)]\ \{\ W_{conf}(E_0,E_0)\ |_{\Theta_{p=0}} +$$

$$\sum_{p=1}^{\infty} W_{conf}(E_0,E_p)\ |_{\Theta_{p\leq 0}}\ \} \tag{21}$$

where following eq's .(6) and (7) the one-phonon exothermic probability is

$$W_{conf}(E_0,E_1) = \pi\ \exp(-\varepsilon_R/\hbar\omega)\ 2[q_0\ (q_C - q_0) + 1]^2\ \exp(-\hbar\omega/\varepsilon_R)$$

while the isothermic probability reads

$$W_{conf}(E_0,E_0) = \pi\ \exp(-\varepsilon_R/\hbar\omega)\ q_0^2.$$

The of exothermic-to-isothermic ratio is found to be

$$W_{conf}(E_0,E_1) / W_{conf}(E_0,E_0) = 2[q_C - q_0 + q_0^{-1}]^2\ \exp(-\hbar\omega/\varepsilon_R)$$

which ratio largely exceeds 1 for standart values of the q-coordinates. A weak *descending* temperature dependence of $\Re_{R0}(T)$ is predicted with the thermal slope:

$$d\Re_R(T)_{lowest\ T}/dT = -\nu\ (1/T)\ (\hbar\omega/k_BT)\ \exp(-\hbar\omega/k_BT)\ \{\ W_{conf}(E_0,E_0)|_{\Theta_{p=0}} +$$

$$\sum_{p=1}^{\infty} W_{conf}(E_0,E_p)|_{\Theta_{p\leq 0}}\ \} < 0$$

From the above ratio and the data of Ref. [12] we conclude that the slope of the temperature dependence of the 1-phonon rate can be 200 fold the slope of the isothermic rate. However, the observed low-temperature rate attributed to 1-phonon absorption (*p*=1) is *ascending* rather than descending. The ascending trend of $\Re_R(T)$ develops only after switching on the lowest excited states through terms of the form $\propto \exp(-\hbar\omega/k_BT)$ mainly.

Accounting for the main contribution of 1-phonon processes we describe the lower temperature rate branch as

$\Re_R(T)_{\text{lower T}} = \nu [1 - \exp(-\hbar\omega/k_B T)] \times$

$\{[ W_{\text{conf}}(E_0,E_0) + W_{\text{conf}}(E_1,E_1) \exp(-\hbar\omega/k_B T) ]|_{\Theta=0} + W_{\text{conf}}(E_0,E_1)|_{\Theta 1<0} +$

$[ W_{\text{conf}}(E_1,E_2)|_{\Theta 1<0} + W_{\text{conf}}(E_1,E_0)|_{\Theta 1>0} ] \exp(-\hbar\omega/k_B T) \}$

$= \nu [1 - \exp(-\hbar\omega/k_B T)] \{ [W_{\text{conf}}(E_0,E_0) + W_{\text{conf}}(E_0,E_1)] +$

$\quad [W_{\text{conf}}(E_1,E_1) + W_{\text{conf}}(E_1,E_2) + W_{\text{conf}}(E_1,E_0)] \exp(-\hbar\omega/k_B T) \}$ (22)

holding good at $k_B T \leq \hbar\omega$. Differentiating we get

$d\Re_R(T)_{\text{lower T}}/dT =$

$\quad \nu (-1/T) (\hbar\omega/k_B T) \exp(-\hbar\omega/k_B T) \{ [W_{\text{conf}}(E_0,E_0) + W_{\text{conf}}(E_0,E_1)] -$

$\quad [W_{\text{conf}}(E_1,E_1) + W_{\text{conf}}(E_1,E_2) + W_{\text{conf}}(E_1,E_0)] [1 - 2\exp(-\hbar\omega/k_B T)] \}$

which is nonnegative at $k_B T < \hbar\omega$ for

$W_{\text{conf}}(E_1,E_1) + W_{\text{conf}}(E_1,E_2) + W_{\text{conf}}(E_1,E_0) \geq W_{\text{conf}}(E_0,E_0) + W_{\text{conf}}(E_0,E_1)$

Now, $d\Re_R(T)_{\text{lower T}}/dT$ is maximal at $T_m$ such that

$\exp(-\hbar\omega/k_B T_m) = (1/4)\{1 - [W_{\text{conf}}(E_0,E_0) + W_{\text{conf}}(E_0,E_1)]/$

$\quad\quad\quad [W_{\text{conf}}(E_1,E_1) + W_{\text{conf}}(E_1,E_2) + W_{\text{conf}}(E_1,E_0)] \}$

Clearly, $0 < \exp(-\hbar\omega/k_B T_m) \leq ¼$. Solving for $T_m$ we get for the "*steepest-ascent temperature*":

$T_m = (\hbar\omega/k_B) / \ln ( 4 / \{1 - [W_{\text{conf}}(E_0,E_0) + W_{\text{conf}}(E_0,E_1)] /$

$\quad\quad\quad [W_{\text{conf}}(E_1,E_1) + W_{\text{conf}}(E_1,E_2) + W_{\text{conf}}(E_1,E_0)] \} )$

The highest limit $T_m$ is attained when the denominator is ¼, that is, when

$W_{\text{conf}}(E_0,E_0) + W_{\text{conf}}(E_0,E_1) \ll W_{\text{conf}}(E_1,E_1) + W_{\text{conf}}(E_1,E_2) + W_{\text{conf}}(E_1,E_0)$

At $\hbar\omega = 75$ meV this gives $T_m = 627$ K $= 54$ meV. The lowest limit $T_m \sim 0$ is obtained for

$W_{conf}(E_0,E_0) + W_{conf}(E_0,E_1) \sim W_{conf}(E_1,E_1) + W_{conf}(E_1,E_2) + W_{conf}(E_1,E_0)$

At intermediate temperatures $0 < T < T_m$ the negative exponent in eq. (22) may be linearized roughly by means of

$$\exp(-\hbar\omega / k_B T) \sim \exp(-\hbar\omega / k_B T_m) (T / T_m)$$

Indeed the function on the right-hand side meets the exponential at both $T = 0$ and $T = T_m$. The quality of the fit at intermediate temperatures is the better the smaller $\hbar\omega / k_B T_m$. Now at $\hbar\omega / k_B T \ll 1$,

$\Re_R(T)_{lower\ T} \cong \nu \{ [W_{conf}(E_0,E_0) + W_{conf}(E_0,E_1)] +$

$\quad [W_{conf}(E_1,E_1) + W_{conf}(E_1,E_2) + W_{conf}(E_1,E_0)] \exp(-\hbar\omega / k_B T) \}$

$\sim \nu \{ [W_{conf}(E_0,E_0) + W_{conf}(E_0,E_1)] + [W_{conf}(E_1,E_1) + W_{conf}(E_1,E_2) +$

$\quad W_{conf}(E_1,E_0)] \exp(-\hbar\omega / k_B T_m) (T / T_m) \}$ \hfill (23)

We stress that the term linear in T appears as one approximates for the negative exponential, as shown above. While the physical significance of the latter manipulation should not be overestimated, the imitated linearity holding good only approximately may help explaining otherwise obscure physical conclusions. (See Appendix II for examples of rates derived under steady-state conditions.)

From equations (6) and (7) using Hermite polynomials $H_n(q)$ for $n \leq 3$, we derive $F_{nm}(q_0,q_C)$ to arrive at

$W_{conf}(E_0,E_0) + W_{conf}(E_0,E_1) \cong$

$\quad \pi q_0^2 \{ 1 + 2(q_C - q_0)^2 \exp(-\hbar\omega/\varepsilon_R) \} \exp(-\varepsilon_R/\hbar\omega),$

$W_{conf}(E_1,E_1) + W_{conf}(E_1,E_2) + W_{conf}(E_1,E_0) \cong$

$\quad \pi \{[2q_0(q_C - q_0) - 1 + 2q_C]^2 + \{[q_0 q_C(2(q_C-q_0)^2 - 1) - (q_C - q_0) + 4\xi_C(q_C - q_0)]^2 +$

$\quad [q_0 q_C - 1]^2 \} \exp(-\hbar\omega/\varepsilon_R) \} \exp(-\varepsilon_R/\hbar\omega)$

with $q_C^{(\pm p)} = \sqrt{[(2/\hbar\omega)\varepsilon_C^{(\pm p)}]}$, $\varepsilon_C^{(\pm p)} = (\varepsilon_R + \Theta_{\pm p})^2/4\varepsilon_R$. Approximate $q_C$ values may be derived assuming that $\varepsilon_R \gg |\Theta_{\pm p}|$. On this ground the bracketed superscripts to $q_C^{(\pm p)}$ will be discarded. The obtained expressions are to be inserted into eq. (23).

## 6. Numerical calculations

We shall illustrate the foregoing analytical conclusions using APES (adiabatic potential energy surface) data obtained by fitting the reaction rate formulae to the experimental temperature dependence of $^{14}$C diffusion in α-iron. Calculated quantities are listed in Table I. Traditional relations between configurational coordinates in eqn. (6) are: right-hand well bottom at $\Delta q_0$, well crossover at $q_C = ½ \Delta q_0$, as the origin $q = 0$ is placed at the left-hand well bottom. We remind that $q = \sqrt{(K/\hbar\omega)}\, Q$ is the scaled configurational coordinate. In as much as $\varepsilon_R \gg |\Theta_{\pm p}|$ to within 3%, we use $p$-independent values for $q_C$ throughout. From Table I data we calculate $\sqrt{(K/\hbar\omega)} = 14.859$ Å$^{-1}$, then $\Delta q_0 \equiv \sqrt{(K/\hbar\omega)}\, \Delta Q_0 = 10.966$ and then $q_C = 5.483$. We compute $\exp(-\varepsilon_R/\hbar\omega) = 8.1\times 10^{-27}$ and $\exp(-\hbar\omega/\varepsilon_R) = 0.983$. From these data we get

$$W_{conf}(E_0,E_0) + W_{conf}(E_0,E_1) \cong 1.839\times 10^{-22}$$

$$W_{conf}(E_1,E_1) + W_{conf}(E_1,E_2) + W_{conf}(E_1,E_0) \cong 2.965\times 10^{-19}$$

whose ratio is $6.204\times 10^{-4}$ (A << B). These highly asymmetric coefficients in the rate equation (17) guarantee the highest possible growth rate with T of the low-temperature combined reaction rate $\Re_R(T)_{low\ T}$. Under these same conditions we also calculate the maximum-ascent temperature $T_m = 643.7$ K, very close to its highest possible value.

Accounting for the asymmetry, the low-temperature relative growth rate at $T_m$ is found to be

$$[T d\Re_R(T)_{lower\ T}/dT] / \Re_R(T)_{lower\ T} \big|_{T=T_m} = (\hbar\omega / k_B T_m)$$

**At** $k_B T_m < \hbar\omega$ the relative growth rate exceeds unity. Now from

$$\Re_R(T)_{lower\ T} = \nu\, [W_{conf}(E_1,E_1) + W_{conf}(E_1,E_2) + W_{conf}(E_1,E_0)]\, \exp(-\hbar\omega / k_B T)$$

$$\sim \nu\, [W_{conf}(E_1,E_1) + W_{conf}(E_1,E_2) + W_{conf}(E_1,E_0)]\, \exp(-\hbar\omega / k_B T_m)\, (T / T_m)$$

we obtain for the linear growth coefficient near maximum ascent

$$A = \nu\, [W_{conf}(E_1,E_1) + W_{conf}(E_1,E_2) + W_{conf}(E_1,E_0)]\, \exp(-\hbar\omega / k_B T_m)\, (1 / T_m)$$

$A = 2.11\times 10^{-9}$ s$^{-1}$K$^{-1}$ at $\nu = 1.86\times 10^{13}$ s$^{-1}$ (77 meV) and $k_B T_m = 55$ meV. This is to be compared with the experimental fitting value of $1.59\times 10^{-9}$ s$^{-1}$K$^{-1}$ from Table I.

Finally, the zero-point rate obtains likewise. From eq. (17) at $k_B T \ll \hbar\omega$:

$$\Re_R(0) = \nu\, [W_{conf}(E_0,E_0) + W_{conf}(E_0,E_1)]$$

wherefrom the above data we calculate $\Re_R(0) = 3.421\times 10^{-9}$ s$^{-1}$. The experimental value from Table I is $D(0)_{th} / P = 9.492\times 10^{-9}$ s$^{-1}$.

## 7. Conclusion

We presented a reaction rate approach to the multiphonon tunneling relaxation in solids. In it, the absorption of *p* phonons during the relaxation is regarded as an exothermic rate process with a zero-point reaction heat amounting to *p* coupled-mode quanta. Alternatively, the emission of *p* phonons during the relaxation is regarded as an endothermic process with a reaction heat amounting to *p* coupled-mode quanta. We analyzed the low-temperature rates predicted on these premises to find that the rate-$\propto$-T signature of the 1-phonon processes holds good only approximately within a temperature range near the maximum ascent temperature of that rate.

In order to assess the proportionality law independently, we note that from the condition that the 1-phonon emission and absorption rates should be proportional to the respective number of phonons $\Delta n_{\Delta E} = [\exp(\Delta E/kT) - 1]^{-1}$ the rates are found to be $k_{em} \propto \Delta E (\Delta n_{\Delta E} + 1)$, $k_{abs} \propto \Delta E \, \Delta n_{\Delta E}$, both of them being $\propto T$ for $kT \gg \Delta E$.

We present numerical calculations to check our reaction rate assignments. For that purpose we took up a typical case of rate versus temperature dependence which increases slowly but steadily within the low temperature tunneling range prior to the transition to the steep classical range. We find the theoretical 1-phonon linear temperature coefficient agreeable to within 30 % with the observed value.

An analysis made indicated that the transition (tunneling) probabilities expanded into Franck-Condon factors to stress the close interconnection between reaction- and multiphonon- relaxation rates.

Acknowledgement. The authors are greatly indebted to the late Professor S.G. Christov for introducing them to the field of Chemical Reaction Rates as well as for his long standing interest and support. It is a privilege to dedicate this paper to Professor Danail Bonchev's 65[th] anniversary.

Appendix 1

Transition Matrix Element

Following Christov [12,13], we define the Hamiltonian to the left of the scaled (dimensionless) crossover coordinate $\xi_C$ as

$$H = H_1 + V^{'}, \qquad \xi \le \xi_C \qquad (AI.1)$$

where $V^{'}$ is the electronic states coupling term finite within a narrow region around $\xi_C$ only. This makes it possible to introduce an auxiliary Hamiltonian

$$H_e \equiv H_1 = \tfrac{1}{2} h\nu \, (\xi^2 - \partial^2/\partial\xi^2), \qquad \xi < \xi_C$$

$$H_e \equiv H_2 = \tfrac{1}{2} h\nu \, [(\xi-\xi_0)^2 - \partial^2/\partial\xi^2] + Q, \qquad \xi > \xi_C \qquad (A1.2)$$

where the origin at the $\xi$–axis is set at the bottom of the left-hand well. We write down Schrödinger's equations for $\xi < \xi_C$ and $\xi > \xi_C$, respectively

$$H_1 \Psi_1 = E_1 \Psi_1$$

$$H_2 \Psi_2 = E_2 \Psi_2 \qquad (A1.3)$$

Having set the above background we define the transition matrix element

$$M_{12} = \langle \Psi_2 | H - H_1 | \Psi_1 \rangle \equiv \langle \Psi_2 | V^{'} | \Psi_1 \rangle \qquad (A1.4)$$

In as much as

$$H_e \Psi_1 = (H_1 + V^{'}) \Psi_1 \qquad (A1.5)$$

We get therefore a system of equations

$$(H_e - H_1)\Psi_1 = (H_1 + V^{'})\Psi_1 - H_1 \Psi_1$$

$$(H_e - H_2)\Psi_2 = (H_1 + V^{'})\Psi_2 - H_2 \Psi_2 \qquad (A1.6)$$

Multiplying the first one by $\Psi_2$ and the second one by $\Psi_1$ to the left and subtracting we obtain using (AI.2)

$$\Psi_2(H_e - H_1)\Psi_1 - \Psi_1(H_e - H_2)\Psi_2 = \Psi_2 H_1 \Psi_1 - \Psi_1 H_1 \Psi_2 + (E_2 - E_1)\Psi_2 \Psi_1 =$$

$$- \tfrac{1}{2} h\nu \, [\Psi_2 (\partial^2 \Psi_1/\partial\xi^2) - \Psi_1(\partial^2 \Psi_1/\partial\xi^2)] + (E_2 - E_1)\Psi_2 \Psi_1 \qquad (A1.7)$$

The latter result is inserted into

$$M_{12} = {}_{-\infty}\!\int^{+\infty} \Psi_2(H_e - H_1)\Psi_1 d\xi \tag{A1.8}$$

which transforms to

$$M_{12} = {}_{\xi_c}\!\int^{+\infty} [\Psi_2(H_e - H_1)\Psi_1 - \Psi_1(H_e - H_2)\Psi_2]\, d\xi \tag{A1.9}$$

because $H_e - H_1 = 0$ ($\xi < \xi_C$) and $(H_e - H_2 = 0$ ($\xi > \xi_C$). Therefore

$$M_{12} = -\tfrac{1}{2}\, h\nu\, {}_{\xi_c}\!\int^{+\infty} [\Psi_2\,(\partial^2\Psi_1/\partial\xi^2) - \Psi_1(\partial^2\Psi_1/\partial\xi^2)]\, d\xi +$$

$$(E_2 - E_1)\, {}_{-\infty}\!\int^{+\infty} \Psi_2 \Psi_1\, d\xi \tag{A1.10}$$

Integrating by parts in the former line and using ground-state functions in the latter line, we get

$$M_{12} = -\tfrac{1}{2}\, h\nu\, [\Psi_2\,(\partial\Psi_1/\partial\xi) - \Psi_1(\partial\Psi_1/\partial\xi)]_{\xi=\xi_c} + (E_2 - E_1)\sqrt{\pi}\, \exp(-2E_{JT}/\, h\nu) \tag{A1.11}$$

in which $E_{JT} = G^2/2K$ is the vibronic coupling energy. The exponent in the latter term is Holstein's reduction factor which makes that term negligible even at different eigenenergies (finite $E_2 - E_1 = \pm p\hbar\omega$). Eq. (A1.11) leads to eq's (1)&(2) at $p = 0$ and to eq's (4)&(5) at $p \neq 0$ for $2E_{LR} \gg h\nu$ which is the usual occurrence. Using harmonic-oscillator wave functions as in eq. (3), the main $\Psi$-term of (A1.11) leads to equations (7) or (7'). (Note that an early misprint in the second term of eq's (7) & (7') has now been corrected [14,15].)

Appendix 2

Steady state rates

The traditional definitions of both reaction and multiphonon rates are based on the energy levels of a vibronic system in which the initial and final electronic states are decoupled. Actually, however, the two constituents do couple so as to secure the vibronic transition. The coupling being provided by both electronic and vibronic means transfigures the vibronic energy levels in the initial and final electronic states. Thus, due to the vibronic tunneling interaction the vibronic energy levels split, the split off levels constituting the energy levels of the coupled system in steady state. The question arises as to whether the rate should be defined under decoupled or coupled conditions.

In migrational rate vs. temperature dependences, the "flat plateau" at low temparatures is in fact slowly increasing $\propto T$ as the temperature is raised. It is also important that the increment is readily observed well before the first excited state of the system starts filling up. While the flat character is intrinsic of the elastic-tunneling reaction rate, the monotonic rise is due to a inelastic-tunneling multiphonon process complementary to the basic isothermic process. The main effect being that of the 1-phonon absorption, we reproduce the 1-phonon rate derived by the Golden Rule for a phonon-coupled two-level system in glasses to represent the rate of an inelastic process [14]:

$$\Re_{mv} = C \coth (\Delta / k_B T ) \qquad\qquad\qquad\qquad\qquad\qquad\qquad\qquad (A2.1)$$

where $\Delta$ is the ground state vibronic tunneling splitting. Eq. (A2.1) reduces to $\Re_{mv} = C(k_B / \Delta) T \equiv AT$ at $T \gg \Delta / k_B$. For a typical value of $\Delta = 0.001$ eV, this gives $T \gg 10$ K to cover most of the plateau range. The T-proportional rate is regarded as the typical signature of the 1-phonon process.

The appearance of the tunneling splitting $\Delta$ distinguishes the steady-state 1-phonon rate (AII.1) from both the traditional reaction rate (10) and the conventional multiphonon rate of eq. (15). An example of a steady-state reaction rate is provided by the relaxation rate of a 2D planar rotator computed by means of the exact Mathieu eigenfunctions of off-center reorientation. Details can be found elsewhere [15].

TABLE 1. Fitting and derivative parameters for the adiabatic potential energy surface controlling carbon diffusion in α-iron (from Ref. [10])

| Fitting parameters | | | | Derivative parameters | | | | |
|---|---|---|---|---|---|---|---|---|
| $\varepsilon_B$ | $\hbar\omega$ | $\eta$ | K | G | $\varepsilon_{CE}$ | $\varepsilon_R$ | $\varepsilon_C$ | $\Delta Q$ |
| [eV] | [eV] | | [eV/Å$^2$] | [eV/Å] | [eV] | [eV] | [eV] | [Å] |
| 1.1 | 0.077 | 0.025 | 17 | 6.272 | 1.157 | 4.626 | 1.158 | 0.738 |

| $2V_{12}$ | $D(0)_{th}{}^{\dagger}$ | $D(0)_{exp}$ | d | $P = f(\zeta/6)\,d^2$ | $f$ | A | | |
|---|---|---|---|---|---|---|---|---|
| [eV] | [cm$^2$/s] | [cm$^2$/s] | [Å] | [cm$^2$] | | [s$^{-1}$K$^{-1}$] | | |
| 0.116 | 3.7×10$^{-27}$<br>1.2×10$^{-24}$ | 6.17×10$^{-25}$ | 0.74 | 6.50×10$^{-17}$ | 1.79 | 1.59×10$^{-9}$ | | |

$^{\dagger}D(0)_{thh} = P\,(\varepsilon_R / \hbar)\,\exp(-\varepsilon_R / \hbar\omega)$
$D(0)_{thv} = P\,(\Delta / k_B)\,A$